\documentclass[sigconf]{acmart}

\usepackage{amssymb}          
\usepackage[noend]{algorithmic}
\usepackage{algorithm} 
\usepackage{multirow}         
\usepackage[table]{xcolor} 
\usepackage{colortbl}         
\usepackage{float}          
\usepackage{subfigure}        
\usepackage{xcolor}
\usepackage{enumitem}

\setlist[itemize]{leftmargin=*}

\AtBeginDocument{
  }

\def\new{\color{black}}

\copyrightyear{2025}
\acmYear{2025}
\setcopyright{acmlicensed}
\acmConference[CIKM '25]{Proceedings of the 34th ACM International Conference on Information and Knowledge Management}{November 10--14, 2025}{Seoul, Republic of Korea}
\acmBooktitle{Proceedings of the 34th ACM International Conference on Information and Knowledge Management (CIKM '25), November 10--14, 2025, Seoul, Republic of Korea}
\acmDOI{10.1145/3746252.3761155}
\acmISBN{979-8-4007-2040-6/2025/11}

\begin{document}

\title{Causality-aware Graph Aggregation Weight Estimator for Popularity Debiasing in Top-K Recommendation}

\author{Yue Que}
\email{yueque2-c@my.cityu.edu.hk}
\orcid{0009-0001-0468-6631}
\affiliation{
  \institution{City University of Hong Kong}
  \city{Hong Kong}
  \country{China}
}

\author{Yingyi Zhang}
\email{yzhang6375-c@my.cityu.edu.hk}
\orcid{0000-0001-9062-3428}
\affiliation{
  \institution{City University of Hong Kong}
  \city{Hong Kong}
  \country{China}
}
\affiliation{
  \institution{Dalian University of Technology}
  \city{Dalian}
  \state{Liaoning}
  \country{China}
}

\author{Xiangyu Zhao}
\authornote{Corresponding author}
\email{xianzhao@cityu.edu.hk}
\orcid{0000-0003-2926-4416}
\affiliation{
  \institution{City University of Hong Kong}
  \city{Hong Kong}
  \country{China}
}

\author{Chen Ma}
\email{chenma@cityu.edu.hk}
\orcid{0000-0001-7933-9813}
\affiliation{
  \institution{City University of Hong Kong}
  \city{Hong Kong}
  \country{China}
}

\renewcommand{\shortauthors}{Yue Que, Yingyi Zhang, Xiangyu Zhao, and Chen Ma}

\begin{abstract}
Graph-based recommender systems leverage neighborhood aggregation to generate node representations, which is highly sensitive to popularity bias, resulting in an echo effect during information propagation.
Existing graph-based debiasing solutions refine the aggregation process with attempts such as edge reconstruction or weight adjustment.
However, these methods remain inadequate in fully alleviating popularity bias.
Specifically, this is because 1) they provide no insights into graph aggregation rationality, thus lacking an optimality guarantee; 2) they fail to well balance the training and debiasing process, which undermines the effectiveness.

In this paper, we propose a novel approach to mitigate popularity bias through rational modeling of the graph aggregation process.
We reveal that graph aggregation is a special form of backdoor adjustment in causal inference, where the aggregation weight corresponds to the historical interaction likelihood distribution.
Based on this insight, we devise an encoder-decoder architecture, namely \textbf{C}ausality-\textbf{a}ware \textbf{G}raph Aggregation Weight \textbf{E}stimator for \textbf{D}ebiasing (CAGED), to approximate the unbiased aggregation weight by optimizing the evidence lower bound of the interaction likelihood.
In order to enhance the debiasing effectiveness during early training stages, we further design a momentum update strategy that incrementally refines the aggregation weight matrix.
Extensive experiments on three datasets demonstrate that CAGED outperforms existing graph-based debiasing methods.
Our implementation is available at \url{https://github.com/QueYork/CAGED}.
\end{abstract}




\maketitle

\begin{figure}[!t]
  \centering
  \includegraphics[width=0.74\linewidth]{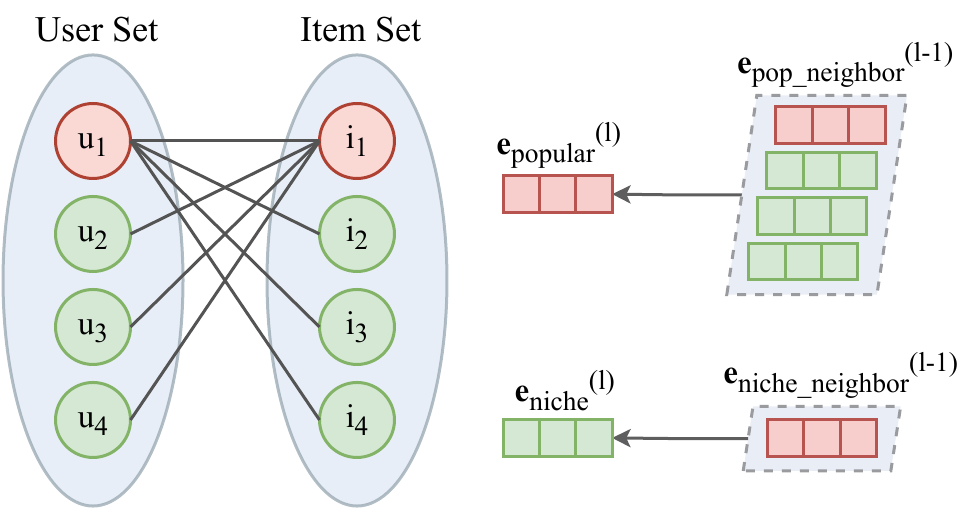}
  \caption{A toy example of popularity bias. Nodes $u_1$, $i_1$ are popular while the rest are niche. In the aggregation process of this bipartite graph (left), popular nodes aggregate the entire opposite set while niche nodes can only incorporate their common popular neighbor (right).}
  \Description{Bipartite graph}
  \label{fig:bipartite}
\end{figure}

\section{Introduction}
Collaborative filtering (CF) is a classic framework in recommender systems to identify users' new interests based on their historical interactions \cite{cf, cf2}.
Among various CF techniques, Graph Convolutional Network (GCN) has emerged as a leading paradigm \cite{gnn}.
By modeling user-item interactions in a bipartite graph and utilizing layer-wise neighborhood aggregation to produce node representations, it has achieved the state-of-the-art performance in Top-K recommendation tasks \cite{lightgcn, ngcf, dgcf}.

Unfortunately, GCNs are highly susceptible to popularity bias, the phenomenon where a small subset of items receives the majority of user feedback \cite{popularitybias}. 
As illustrated in Figure \ref{fig:bipartite}, popularity bias manifests as extreme imbalance in node degrees.
The presence of such degree imbalance causes high-degree (popular) nodes to dominate the information propagation, since less-connected (niche) nodes can only aggregate embeddings from popular nodes \cite{degree}.
Following this, node representations risk converging to similar values through graph aggregation, thereby impairing their representativeness and discriminative capability.
Consequently, GCN models require proper popularity inhibition in the interaction graph.

Existing graph-based debiasing solutions mainly concentrate on refining two aspects of the aggregation process.
\textbf{1) Edge reconstruction:} this method involves strategically augmenting or dropping specific interaction edges to construct a bias-free user-item graph \cite{galore, advdrop, EdgeClassification, dropedge}.
\textbf{2) Aggregation re-weighting:} this idea adjusts edge weights in graph aggregation to balance the proportion of mixing popular node representations into niche nodes \cite{adjnorm, apda}.  

{\new However, we argue that these graph-based methods have two irrationalities that keep them from fully addressing popularity bias.
\textbf{1) Irrationality in graph aggregation}, the limitation that current methods lack exploration in graph aggregation rationality.
Popularity bias amplifies the representation of popular items within the graph, while GCNs operate as black boxes, leaving the rationale in obtaining recommendations from graph aggregation untouched \cite{Causalsurvey}.
On this basis, when introducing further refinements, existing approaches simply take the aggregation mechanism for granted, which hinders both interpretability and the ability to build theoretically grounded effectiveness.
\textbf{2) Irrationality in balancing training and debiasing}, the limitation that current methods exploit under-trained representations for debiasing. 
Graph-based approaches commonly utilize node representations to capture graph information, while ignoring representation qualities at different training stages. 
Therefore, during early training stages, under-trained representations would introduce additional noise into bias mitigation and impair debiasing effectiveness.
To better address popularity bias, it is essential to demystify the recommendation rationale from graph aggregation and also reasonably balance the reliance on under-trained representations.} 

{\new To conceptualize the rationale behind recommendations, we leverage causality theory to instantiate popularity bias as a confounder \cite{Causalsurvey}. 
In causal inference, confounders are addressed by techniques such as backdoor adjustment, which explicitly models their indirect causal effects to the target and incorporates them into the total estimation \cite{causalinference}.
However, in recommendation scenario, the confounder representing popularity bias is typically endogenous, because it arises from closed-loop feedback where users are exposed only to items recommended by the deployed system, skewing user interactions toward algorithm-favored items \cite{longtail, closedloop2, closedloop3}.
As a result, only the biased feedback collection is observable, while underlying distribution of the confounding effect keeps inaccessible.
To address this challenge, we introduce variational inference to approximate this non-closed distribution by alternatively finding its evidence lower bound (ELBO) \cite{VariationalInference}.
We extend this workflow not only to explain the recommendation rationale in graph aggregation, but also to eventually construct the unbiased user-item graph.
}

Hence, in this paper, we propose a novel framework to mitigate popularity bias.
\textbf{First}, to solve the irrationality in graph aggregation, we construct a general causal graph to capture causal relations among the confounder, user nodes, and recommendation results. 
We then substantiate that graph aggregation in fact operates as backdoor adjustment on the aforementioned causal graph, where edge weights correspond to past interaction likelihoods.
On this basis, we devise our model to learn the unbiased weights by estimating actual history likelihoods, namely \underline{C}ausality-\underline{a}ware \underline{G}raph Aggregation Weight \underline{E}stimator for \underline{D}ebiasing (CAGED), which employs an encoder-decoder architecture taking inspiration from Variational Auto-Encoder (VAE) \cite{VAE, cvae, cfvae}.
\textbf{Second}, to solve the irrationality in balancing training and debiasing, we propose the momentum update strategy to train CAGED, which adjusts debiasing to incrementally take effective with sufficient representation training.
Empirical experiments conducted on three real-world datasets with the LightGCN backbone \cite{lightgcn} show that CAGED outperforms existing graph-based debiasing approaches.

In summary, our contributions are threefold.
\begin{itemize}
    \item {\new We introduce causal modeling of graph aggregation to uncover its recommendation rationale. By abstracting it as a form of backdoor adjustment, we address the irrationality in graph aggregation.}
    \item We devise the CAGED method to learn graph aggregation weights. {\new More importantly, we propose the momentum update strategy to address the irrationality in balancing training and debiasing.}
    \item We conduct comprehensive experiments on three real-world datasets, demonstrating the superiority of CAGED over existing baselines in both recommendation and debiasing abilities.
\end{itemize}
\section{Preliminaries}
We first outline the graph aggregation in GCN recommender models, then we introduce the backdoor adjustment in causal inference.

\subsection{Graph Aggregation}
Graph-based recommendation systems utilize a bipartite graph structure to represent user-item interactions, $\mathcal{G}=(\{\mathcal{U} \cup \mathcal{I}\}, \mathcal{E})$, where $\mathcal{U}$ is the user set with $|\mathcal{U}| = M$, $\mathcal{I}$ is the item set with $|\mathcal{I}| = N$, and $\mathcal{E}$ is the user-item edge set.

Taking LightGCN as the paradigm, graph aggregation is a layer-wise operation to generate stacked embeddings for each $u \in \mathcal{U}$ and $i \in \mathcal{I}$.
The initial embeddings for $u$ and $i$ are $\mathbf{e}_u^{(0)}, \mathbf{e}_i^{(0)} \in \mathbb{R}^\mathcal{K}$, where $\mathcal{K}$ denotes the hidden dimensionality. $N(x)$ refers to $x$'s neighbor set. At each layer $l = 1, ..., L$, graph aggregation is defined as:
\begin{align}
    \mathbf{e}_u^{(l)} &= \sum_{i \in \mathcal{N}(u)} \frac{1}{\sqrt{|\mathcal{N}(u)| |\mathcal{N}(i)|}} \mathbf{e}_i^{(l-1)}, \\
    \mathbf{e}_i^{(l)} &= \sum_{u \in \mathcal{N}(i)} \frac{1}{\sqrt{|\mathcal{N}(i)| |\mathcal{N}(u)|}} \mathbf{e}_u^{(l-1)}.
\end{align}

We write this process alternatively in a matrix form. 
Let the adjacency matrix of $\mathcal{G}$ be $\mathbf{A} \in \mathbb{R}^{(M+N) \times (M+N)}$ and the diagonal degree matrix be $\mathbf{D} \in \mathbb{R}^{(M+N) \times (M+N)}$.
Denoting the embedding matrix as $\mathbf{E} \in \mathbb{R}^{(M+N) \times \mathcal{K}}$, we have:
\begin{equation}
    \mathbf{E}^{(l)} = \mathbf{D}^{-\frac{1}{2}} \mathbf{A} \mathbf{D}^{-\frac{1}{2}} \mathbf{E}^{(l-1)} = \widetilde{\mathbf{A}} \mathbf{E}^{(l-1)},
\end{equation}
where $\widetilde{\mathbf{A}}$ is derived to represent the aggregation weights.

After $L$ layers of aggregation, LightGCN adopts the average pooling operation to produce the final representations:
\begin{equation}
    \mathbf{E} = \frac{1}{L+1} \sum_{l=0}^L \mathbf{E}^{(l)}.
\end{equation}

\subsection{Backdoor Adjustment}

\begin{figure}[t]
    \centering
    \subfigure[]{
        \includegraphics[width=0.42\linewidth]{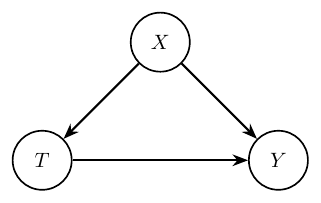}
        \label{fig:causalexample}
    }
    \hfill
    \subfigure[]{
        \includegraphics[width=0.42\linewidth]{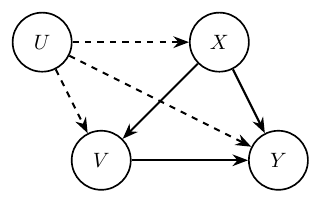}
        \label{fig:rsgraph}
    }
    \caption{(a): A typical causal graph example. (b): Causal graph of general recommendation scenario.}
    \Description{Causal graphs}
    \label{fig:causalgraphs}
\end{figure}

Causal graph is a directed acyclic graph where edges are directed from causes to effects.
Take Figure \ref{fig:causalexample} as an example, $T$ represents the treatment variable and $Y$ represents the outcome variable. 
$X$ is identified as a confounder because it affects $Y$ via the path $T \rightarrow Y$ and also directly influences $Y$, interfering the association between $T$ and $Y$. 
Formally speaking, we say $X$ satisfies the backdoor criterion with respect to $(T, Y)$ for holding two properties \cite{causaltextbook}: (1) $X$ is not a descendant of $T$; (2) $X$ blocks every path between $T$ and $Y$ that contains an arrow pointing into $T$.


To accurately estimate the causal effect of $T$ on $Y$, intervention is employed to fix the treatment variable to a specific value $t$ through the $do(T=t)$ operation, which removes any confounding factors that may influence the natural distribution of $T$.
The deconfounded causal effect is then expressed as $P(Y \mid do(T=t))$.
In other words, intervention isolates the causal path $T \rightarrow Y$ by disabling alternative pathways $X \rightarrow T$ and $X \rightarrow Y$ to eliminate confounder $X$.
When the confounder satisfies the backdoor criterion, backdoor adjustment is an effective way for implementing intervention by conditioning on the confounder.
This allows for an unbiased estimation of the causal effect, expressed as follows:
\begin{equation}
    P(Y \mid do(T=t)) = \sum_{x} P(Y \mid T=t, X=x) P(X=x).
\end{equation}

\section{Methodology}
We first present the causal modeling to graph aggregation, which interprets it as estimating the causal effect of recommendations on users using backdoor adjustment.
Next, we demonstrate CAGED, an encoder-decoder model for learning unbiased aggregation weights.
Finally, we introduce the momentum update strategy, which incrementally integrates these weights to ensure effective debiasing.

\subsection{Causal Explanation of Graph Aggregation}
As illustrated in Figure \ref{fig:rsgraph}, we construct a causal graph to represent the typical recommendation scenario. 
We identify 4 key components from this scenario, denoted as the graph nodes:
\begin{itemize}
    \item \textit{Node} ($U$): any user (node).
    \item \textit{History} ($X$): the record of $U$'s past interacted items (nodes).
    \item \textit{Intervention} ($V$): the candidate item (node) recommended by the system to interact with $U$.
    \item \textit{Preference} ($Y$): the extent that $U$ favors the intervention.
\end{itemize}
The causal relations among these nodes are labeled as directed edges. 
The meaning of each edge is outlined as follows:
\begin{itemize}
    \item $U \rightarrow (X, V, Y)$: these relations labeled in dashed lines merely mean $X$, $V$, and $Y$ are affiliated with $U$.
    \item $X \rightarrow V$: this relation explains how historical interactions influence the intervention. Specifically, CF model learns from historical data to determine new recommendations.
    \item $X \rightarrow Y$: this relation suggests that history reflects preferences.
    \item $V \rightarrow Y$: this relation captures the feedback to the intervention.
\end{itemize}

{\new To perform causal inference in this scenario, the intervention is defined as recommending item $v$ to user $u$, and the target is to estimate $u$'s preference on the given $v$, that is, $P(Y=y \mid U=u, do(V=v))$.
In the causal graph, $X$, satisfying the backdoor criterion, serves as the confounder.
Since $X$ actually reflects $U$'s popularity distribution, deconfounding $X$'s causal effect effectively corresponds to populairty debiasing.
Therefore, we apply backdoor adjustment to factorize the intervention probability for each $X=x$, which is:
\begin{equation}
\begin{aligned}
    P(Y=y \mid U=u, do(V=v)) &= \sum_{x} p(y | u, v, x) p(x | u) \\
    & = \mathbb{E}_{x \sim p(x|u)} \left[ p(y|u,v,x)\right].
\end{aligned}
\label{eq:ba}
\end{equation}

We assume: 1) user $u$'s preference for an intervention $v$ in GCN is represented as the inner product of embeddings $\mathbf{e}_u$ and $\mathbf{e}_v$, \textit{i.e.}, $p(y|u, do(v)) = \mathbf{e}_u^T\mathbf{e}_v$; 2) user $u$'s each history interaction $x|u$ is conditionally independent to each other.
We define a normalizing function $F(u)$ w.r.t to user $u$ as:
\begin{equation}
    F(u) = \sum_{x \in \mathcal{N}(u)} \frac{1}{\sqrt{|\mathcal{N}(u)| |\mathcal{N}(x)|}}.
\end{equation}
Similar to backdoor adjustment, one-layer aggregation of $u$ at layer $l$ also follows the expectation form under $p(x|u)$ distribution: 
\begin{equation}
\begin{aligned}
    \mathbf{e}_u^{(l)} &= \sum_{x \in \mathcal{N}(u)} \frac{1}{\sqrt{|\mathcal{N}(u)| |\mathcal{N}(x)|}} \mathbf{e}_x^{(l-1)} \\
    &= \sum_{x \in \mathcal{N}(u)} \frac{F(u)}{F(u)\sqrt{|\mathcal{N}(u)| |\mathcal{N}(x)|}} \mathbf{e}_x^{(l-1)} \\
    &= \mathbb{E}_{x \sim p(x|u)} \left[F(u)\mathbf{e}_x^{(l-1)}\right], \\
    \text{where } &p(x|u) = \frac{1}{F(u)\sqrt{|\mathcal{N}(u)| |\mathcal{N}(x)|}} = \frac{1/\sqrt{|\mathcal{N}(x)|}}{\sum_{x} 1/\sqrt{|\mathcal{N}(x)|}}.
\end{aligned}
\label{eq:eq}
\end{equation}
In $p(x|u)$ derivation, $F(u)$ is introduced to normalize it into a valid probability distribution satisfying $\sum_xp(x|u)=1$.
By stacking multiple such expectations from each layer to form the final embedding, we can naturally write the inner product between $u$ and $v$ into the same expectation form:
\begin{equation}
\begin{aligned}
    \mathbf{e}_u^T\mathbf{e}_v &= \left[ \frac{1}{L+1} \left( \mathbf{e}_u^{(0)} + \sum_{l=1}^{L}\mathbb{E}_{x \sim p(x|u)} \left[F(u)\mathbf{e}_x^{(l-1)}\right] \right) \right]^T\mathbf{e}_v \\
    &=\mathbb{E}_{x \sim p(x|u)} \left[ \frac{1}{L+1}\left( \mathbf{e}_u^{(0)} + \sum_{l=1}^{L}F(u)\mathbf{e}_x^{(l-1)} \right) ^T\mathbf{e}_v \right].
\end{aligned}
\label{eq:ga}
\end{equation}

At last, Eq. (\ref{eq:ba}) and (\ref{eq:ga}) demonstrate that, under our predefined assumptions, graph aggregation operates as a specialized form of backdoor adjustment: it maps $p(x|u)$ to the inverse square root degree ratio and instantiates $p(y|u,v,x)$ as a combination of the corresponding embeddings.
Moreover, we find graph aggregation weight equals to $F(u)\cdot p(x|u)$.
However, as previously discussed, $p(x|u)$ actually conforms to a non-closed form. 
GCNs' simplistic modeling of $p(x|u)$ in Eq. (\ref{eq:eq}) fails to reasonably estimate the causal effect from confounder $X$, leaving graph-based models consistently vulnerable to popularity bias.
To overcome this limitation, we devise the CAGED model to learn actual historical likelihood distributions, then apply them back to GCN as the unbiased aggregation weights.
}

\begin{figure*}[t]  
    \centering  
    \includegraphics[width=\textwidth]{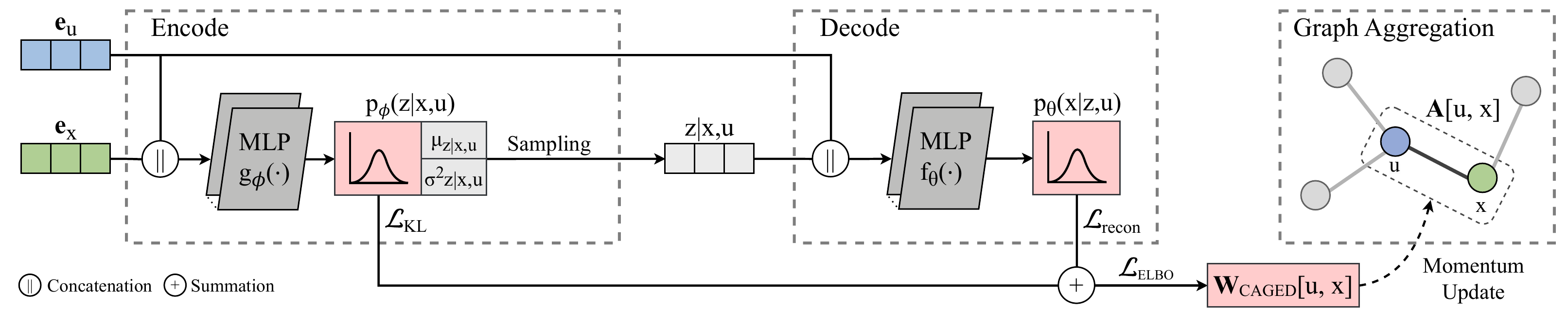}  
    \caption{An illustration of CAGED architecture. Embeddings of the convolution center $u$ and one of its neighbors $x$ are forwarded into the encoder and decoder modules respectively. The ELBO output from two modules forms the CAGED-generated weight $\mathbf{W_{CAGED}[u, x]}$ to incrementally update the original aggregation weight $\mathbf{A[u, x]}$.}
    \label{fig:model}  
        \Description{model}
\end{figure*}
\subsection{CAGED}
Following the above analysis, we propose CAGED to model $p(x | u)$.
However, directly accessing the ground-truth distribution of $X$ conditioned on $U$, denoted as $p_\theta(x | u)$, is impractical, because we are in lack of prior knowledge about the collection process of past interactions.
{\new Thus, we draw inspiration from variational inference to explicitly model the collection process as a set of feasible distributions conditioned on each node, from which interactions could be sampled to compute likelihoods.} 
To achieve this, we introduce a latent variable $Z$ (\textit{cf.} Figure \ref{fig:model}) to represent the deployed system's characteristics that shape the closed-loop feedback. 
{\new The feasible distributions of $Z$ are denoted as $p_\phi(z | u)$, from which the unknown $p_\theta(x | u)$ is assumed to be sampled. 
Consequently, $p_\phi(z|x,u)$ establishes an evidence lower bound (ELBO) for the non-closed $p_\theta(x | u)$, enabling us to approximate the true likelihood by raising the ELBO toward its upper limit.}

\subsubsection{\textbf{Derivation of the ELBO}}
We assume the prior of $z|u$ follows a standard Gaussian, \textit{i.e.}, $p_\theta(z|u)\sim \mathcal{N}(0,\mathbf{I})$. 
Meanwhile, the marginal distribution $z|x,u$ follows a Gaussian slightly deviating from the standard one to capture the unique characteristics of $x$, \textit{i.e.}, $p_\phi(z|x,u)\sim \mathcal{N}(\boldsymbol{\mu}_{z|x,u}, \boldsymbol{\sigma}^2_{z|x,u}\mathbf{I})$. 
With these assumptions, we proceed to calculate the log-likelihood to produce the ELBO:
\begin{equation}
\begin{aligned}
&\log p_\theta(x|u) = \log \int p_\phi(z|x,u) \frac{p_\theta(x,z|u)}{p_\phi(z|x,u)} \, d(z|u) \\
&\geq \int p_\phi(z|x,u) \log \frac{p_\theta(x|z,u)p_\theta(z|u)}{p_\phi(z|x,u)} \, d(z|u) \; \; \text{\small(Jensen's Inequality)} \\ 
&= \mathbb{E}_{p_\phi(z|x,u)}\left[\log p_\theta(x|z,u)\right] - \mathbb{E}_{p_\phi(z|x,u)}\left[ \log \frac{p_\phi(z|x,u)}{p_\theta(z|u)} \right] \\
&= \mathbb{E}_{p_\phi(z|x,u)}\left[\log p_\theta(x|z,u)\right] - KL\left(p_\phi(z|x,u)\,||\,p_\theta(z|u)\right). \\
\end{aligned}
\end{equation}
The ELBO consists of two components. 
$\mathbb{E}_{p_\phi(z|x,u)}\left[\log p_\theta(x|z,u)\right]$ is the reconstruction term that estimates the expected likelihood of $x|u$ given $z$ under the feasible posterior $p_\phi(z|x,u)$. 
The second term $KL\left(p_\phi(z|x,u)\,||\,p_\theta(z|u)\right)$ is the Kullback-Leibler (KL) divergence to regularize the difference between the marginal posterior and the prior, ensuring a sound sampling of latent variable $z|u$. 
Combining them, the ELBO establishes a robust lower bound for $p_\theta(x|u)$, the true distribution of our historical interaction likelihood. 
We could approximate the ground-truth $p_\theta(x|u)$ by optimizing the ELBO. 
Therefore, our optimization problem is formulated as:
\begin{equation}
    \mathop{\min}\limits_{\theta, \phi} -\mathbb{E}_{p_\phi(z|x,u)}\left[\log p_\theta(x|z,u)\right] + KL\left(p_\phi(z|x,u)\, ||\, p_\theta(z|u)\right).
\end{equation}

\subsubsection{\textbf{Model Architecture}}
The objective of CAGED is to solve the above optimization problem.
As shown in Figure \ref{fig:model}, we adopt an encoder-decoder architecture taking inspiration from VAE.
With totally different goals, we directly leverage the ELBO loss to approximate data likelihoods, instead of utilizing it to reconstruct samples for generative purpose.
CAGED processes the embeddings of $x$ and $u$ through the encoder and decoder, outputting $\mathcal{L}_{ELBO}$.
The ELBO result is then encapsulated into $W_{CAGED}$ and fed back into GCN as the aggregation weight, enabling unbiased representation learning.

The encoder module is designed to model $p_\phi(z|x,u)$, which defines a unique latent space for the input nodes $x$ and $u$.  
Given that we have predefined the posterior $p_\phi(z|x,u)$ to follow a Gaussian, the encoder leverages a Multi-Layer Perceptron (MLP) to fit this distribution, outputting its mean $\boldsymbol{\mu}_{z|x,u}$ and variance $\boldsymbol{\sigma}^2_{z|x,u}$.
The ELBO regularization term is then calculated as:
\begin{equation}
\begin{aligned}
    \mathcal{L}_{KL} &= KL\left( \mathcal{N}(\boldsymbol{\mu}_{z|x,u}, \boldsymbol{\sigma}^2_{z|x,u}\mathbf{I}) || \mathcal{N}(0, \mathbf{I}) \right) \\
    &= \beta \sum_{k=1}^\mathcal{K} \frac{1}{2}\left( \mu_k^2 + \sigma_k^2 - \log\sigma_k^2 - 1 \right).
\end{aligned}
\end{equation}
Here, we adopt the setting in $\beta$-VAE \cite{betavae} to put an additional hyper-parameter $\beta$ in $\mathcal{L}_{KL}$. 
It adjusts the scale of KL score to better match the aggregation weight.
Subsequently, each latent variable $z|x,u$ is sampled using the reparameterization trick, ensuring the sampling process remains differentiable:
\begin{equation}
z|x,u = \boldsymbol{\mu}_{z|x,u} + \tau \cdot \boldsymbol{\sigma}_{z|x,u}, \quad \tau \sim \mathcal{N}(0, \mathbf{I}),    
\end{equation}
where $\tau$ is the random noise drawn from a standard Gaussian.

The decoder module in CAGED generates the unbiased likelihood as $p_\theta(x|z,u)$, which incorporates latent information about the collection process to refine the original $p_\theta(x|u)$.
To simplify this process for implementation, $p_\theta(x|z,u)$ is parameterized as another Gaussian.
In this setup, the decoder network simply takes the latent variable $z|x,u$ and the condition $u$ as inputs, and directly outputs the reconstructed $\hat{x}$, which is considered to be the expectation of $p_\theta(x|z,u)$.
Consequently, the Euclidean distance (\textit{i.e.}, mean squared error) between the representations $\mathbf{e}_x$ and $\mathbf{e}_{\hat{x}}$ could be employed to measure the logarithmic probability of sampling $x$ from the unbiased likelihood $p_\theta(x|z,u)$:
\begin{equation}
\begin{aligned}
    &\mathcal{L}_{recon} = -\mathbb{E}_{p_\phi(z|x,u)}\left[\log p_\theta(x|z,u)\right] \\
    &\approx -\log p_\theta(x|z,u) \quad \text{\small(Stochastic sampling for simplicity)} \\
    &= \frac{1}{2\sigma_\theta^2}\| \mathbf{e}_x - \mathbf{e}_{\hat{x}} \|_2^2 + \frac{\mathcal{K}}{2}\log(2\pi\sigma_\theta^2) \\
    &\propto \lambda \| \mathbf{e}_x - \mathbf{e}_{\hat{x}} \|_2^2,
\end{aligned}
\end{equation}
where $\lambda$ controls the variance of $p_\theta(x|z,u)$, serving as the other scaling hyper-parameter when transforming to aggregation weight.

\subsubsection{\textbf{Optimization with LightGCN}}
With the above derivation, the ELBO loss for CAGED training is defined as:
\begin{equation}
\begin{aligned}
    \mathcal{L}_{ELBO} &= \mathcal{L}_{recon} + \mathcal{L}_{KL} \\
    &= \lambda \| \mathbf{e}_x - \mathbf{e}_{\hat{x}} \|_2^2 + \beta \sum_{k=1}^\mathcal{K} \frac{1}{2}\left( \mu_k^2 + \sigma_k^2 - \log\sigma_k^2 - 1 \right).
\end{aligned}    
\end{equation}
After optimization, {\new $F(u) \cdot exp(-\mathcal{L}_{ELBO})$} would be utilized as the new aggregation weight in LightGCN's graph aggregation.

For LightGCN, it leverages the pairwise Bayesian Personalized Ranking (BPR) loss \cite{BPR} to maximize the difference in inner product scores between positive and negative items:
\begin{equation}
    \mathcal{L}_{BPR} = \sum_{u,i \in \mathcal{N}(u), j \notin \mathcal{N}(u)} -\ln \sigma( \mathbf{e}_u^T \mathbf{e}_i - \mathbf{e}_u^T\mathbf{e}_j) + \gamma \| \Theta \|_2^2,
\end{equation}
where $\sigma$ denotes the sigmoid function, and $\gamma$ represents the regularization hyper-parameter.

\subsubsection{\textbf{Two-stage Training}}
Since the representations learned by LightGCN are often insufficiently refined at the initial stage of training,
CAGED is prone to being misled when learning from these low-quality embeddings.
To this end, we adopt a two-stage training within each training epoch to facilitate the co-convergence of both models.
In the first stage, LightGCN experiences regular training to generate more robust representations.
In the second stage, CAGED is trained conditionally only if the first stage improves the performance on the validation data. 
We then update the aggregation weight matrix with the CAGED-generated weights.

\begin{algorithm}[t]
    \renewcommand{\algorithmicrequire}{\textbf{Initialize:}}
	\caption{Momentum update process}
    \label{alg}
    \begin{algorithmic}[1] 
        \REQUIRE default aggregation weight matrix $\widetilde{\mathbf{A}}^{(0)} = \mathbf{D}^{-\frac{1}{2}} \mathbf{A} \mathbf{D}^{-\frac{1}{2}}$, embedding matrix $\mathbf{E}^{(0)}$, and mixture factor $\epsilon$;
        \FOR{epoch $t = 1, \cdots, T$}
            \STATE $\mathbf{E}^{(t)} \leftarrow$ train GCN with $\widetilde{\mathbf{A}}^{(t-1)}$ and $\mathbf{E}^{(t-1)}$;
            \IF{update condition satisfies}
                \STATE $\mathbf{W}_{CAGED} \leftarrow$ train CAGED with $\mathbf{E}^{(t)}$;
                \STATE $\widetilde{\mathbf{A}}^{(t)} = (1-\epsilon)\widetilde{\mathbf{A}}^{(t-1)} + \epsilon\mathbf{W}_{CAGED}$;
            \ELSE
                \STATE $\widetilde{\mathbf{A}}^{(t)} = \widetilde{\mathbf{A}}^{(t-1)}$;
            \ENDIF
            
        \ENDFOR
    \end{algorithmic}
\end{algorithm}

\subsubsection{\textbf{Momentum Update Strategy}}
We name the strategy for updating the aggregation weight matrix as momentum update, because CAGED-generated weights in each update epoch are accumulated into the aggregation weight matrix instead of direct replacement.
As outlined in Algorithm \ref{alg}, the aggregation weight matrix is initialized to the original one used in LightGCN.
At each update step $t$, the likelihood matrix generated by CAGED, $W_{CAGED}$, is gradually integrated into the current weight matrix $\widetilde{\mathbf{A}}^{(t)}$, with the hyper-parameter $\epsilon$ controlling the extent of mixture.


This strategy ensures high-quality embedding initialization for CAGED learning, mitigating the impact of under-trained representations on debiasing effectiveness. 
By accumulating historical updates, it also facilitates smoother weight transitions, enhancing convergence stability and preventing abrupt performance changes.
\section{Experiment}
In this section, we conduct extensive experiments to evaluate the effectiveness of CAGED on Top-K recommendation task.
We aim to answer the following research questions:
\begin{itemize}
    \item \textbf{RQ1:} How does CAGED perform comparing with existing graph-based and model-agnostic debiasing methods in terms of Top-K recommendation and popularity debiasing?
    \item \textbf{RQ2:} Can CAGED significantly debias the item popularity?
    \item \textbf{RQ3:} How does the proposed training strategy affect model performance?
    \item \textbf{RQ4:} How do different components of CAGED contribute to the overall recommendation performance?
\end{itemize}

\subsection{Datasets}
\begin{figure*}[t!]
    \centering  
    \subfigure[MovieLens]{\includegraphics[width=0.23\linewidth]{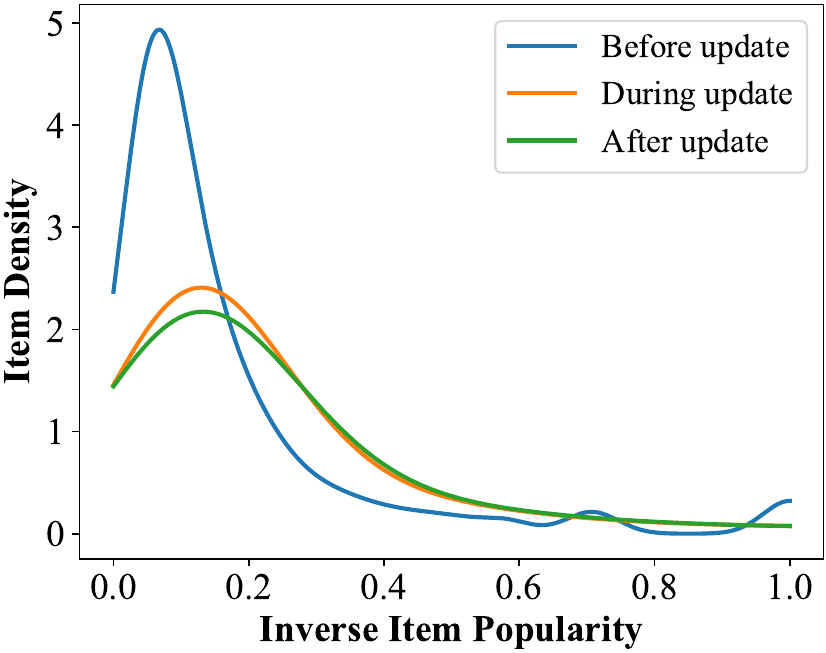}}
    \hspace{3em}
    \subfigure[Pinterest]{\includegraphics[width=0.23\linewidth]{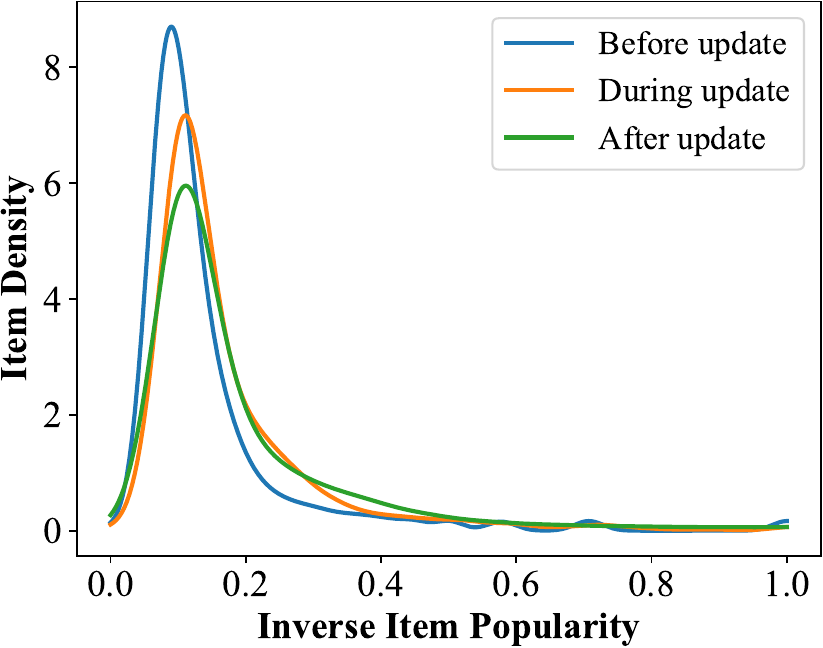}}
    \hspace{3em}
    \subfigure[Epinions]{\includegraphics[width=0.24\linewidth]{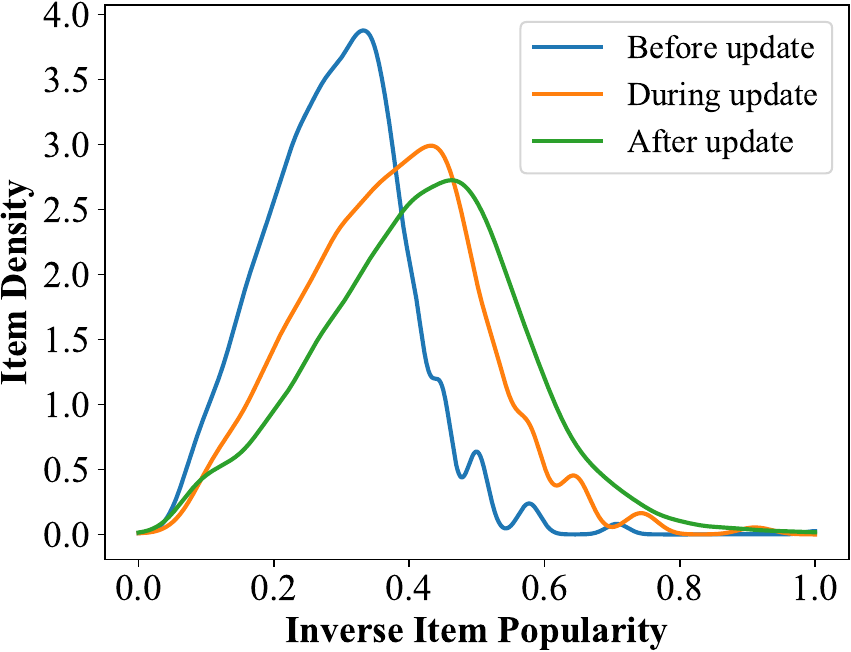}}
    \Description{density}
    \caption{\new Visualization of inverse item popularity before, during, and after CAGED update.}
    \label{fig:density}
\end{figure*}

To ensure a fair comparison, we conduct experiments using three publicly available real-world datasets, with detailed statistics summarized in Table \ref{tab:dataset-stats}.
MovieLens\footnote{\url{https://grouplens.org/datasets/movielens/1m/}} \cite{movielens} is a widely used benchmark that contains explicit movie ratings (1-5 scale). For this study, we employ the MovieLens-1M version.
Pinterest\footnote{\url{https://sites.google.com/site/xueatalphabeta/dataset/pinterest_iccv}} \cite{pinterest} is an image recommendation dataset that records the pins over images initiated by users.
Epinions\footnote{\url{http://www.trustlet.org/epinions.html}} is collected from a product review platform, we adopt its cropped version from the work \cite{apda}.
{\new All selected datasets exhibit a long-tail pattern in inverse item popularity, indicating the edge distribution heavily skewed toward head items, as demonstrated by the "before update" curves in Figure \ref{fig:density}.
This property makes them particularly suitable for evaluating recommendation improvements in the presence of popularity bias.}

\begin{table}[h!]
    \centering
    \caption{Dataset statistics.}
    \label{tab:dataset-stats}
    \begin{tabular}{lcccc}
        \toprule
        \textbf{Dataset} & \textbf{User \#} & \textbf{Item \#} & \textbf{Interaction \#} & \textbf{Density} \\
        \midrule
        MovieLens & 6,040 & 3,952 & 1,000,209 & 0.04190 \\
        Pinterest & 55,186 & 9,916 & 1,463,556 & 0.00267 \\
        Epinions & 11,496 & 11,656 & 327,942 & 0.00245 \\
        \bottomrule
    \end{tabular}
\end{table}

\begin{table*}[t!]
    \centering
    \caption{Performance comparison. The best results are highlighted in boldface and the second-best results are underlined. Gain denotes the performance improvement over vanilla LightGCN. Superscript {\Large *} indicates the statistically significant gain of CAGED (\textit{i.e.}, two-sided t-test with $\mathbf{p<0.05}$).
    }
    \label{tab:main}
    \resizebox{0.999\textwidth}{!}{\new
        \begin{tabular}{c c c c c c c c c c c c c c} 
            \toprule
            \multirow{2}{*}{Dataset} & \multirow{2}{*}{Method} & \multicolumn{4}{c}{All Items} & \multicolumn{4}{c}{Niche Items} & \multicolumn{4}{c}{Popular Items} \\
            \cmidrule(lr){3-6} \cmidrule(lr){7-10} \cmidrule(lr){11-14}
            &&R@20 & Gain & N@20 & Gain & R@20 & Gain & N@20 & Gain & R@20 & Gain & N@20 & Gain \\
            \midrule
            \multirow{8}*{MovieLens}&LightGCN    & 0.2758& N.A.& 0.4803& N.A.& 0.0440& N.A.& 0.0434& N.A.& 0.3605& N.A.& 0.4854& N.A.\\
            &AdvDrop      & 0.2767& +0.33\%& 0.4823& +0.42\%& 0.0472& +7.27\%& 0.0466& +7.37\%& 0.3594& -0.31\%& 0.4847& -0.14\%\\
            &GALORE      & 0.2760& +0.07\%& 0.4811& +0.17\%& 0.0455& +3.41\%& 0.0440& +1.38\%& 0.3609& +0.11\%& 0.4860& +0.12\%\\
            &APDA        & 0.2754& -0.15\%& 0.4787& -0.33\%& 0.0490& +11.36\%& 0.0474& +9.22\%& 0.3571 & -0.94\%& 0.4825 & -0.60\%\\
            &AdjNorm     & \underline{0.2785}& \underline{+0.98\%}& \underline{0.4837}& \underline{+0.71\%}& 0.0430& -2.27\%& 0.0356& -17.97\%& \textbf{0.3656}& \textbf{+1.41\%}& \textbf{0.4959}& \textbf{+2.16\%}\\
            &Reg         & 0.2750& -0.29\%& 0.4793& -0.21\%& 0.0430& -2.27\%& 0.0411& -5.30\%& 0.3617 & +0.33\%& 0.4863 & +0.19\%\\
            &IPS         & 0.2772& +0.51\%& 0.4808& +0.10\%& \underline{0.0511}& \underline{+16.14\%}& \underline{0.0519}& \underline{+19.59\%}& 0.3590& -0.42\%& 0.4824& -0.62\%\\
            &MACR         & 0.2743& -0.54\%& 0.4768& -0.73\%& 0.0450& +2.16\%& 0.0495& +14.06\%& 0.3583& -0.61\%& 0.4778& -1.57\%\\
            &CAGED & \textbf{0.2800}& \textbf{+1.52\%}*& \textbf{0.4852}& \textbf{+1.02\%}*& \textbf{0.0537}& \textbf{+22.05\%}*& \textbf{0.0533}& \textbf{+22.81\%}*& \underline{0.3619}& \underline{+0.39\%}*& \underline{0.4872}& \underline{+0.37\%}*\\
            \midrule
            \multirow{8}*{Pinterest}& LightGCN    & 0.1627& N.A.& 0.1739& N.A.& 0.0715& N.A.& 0.0507& N.A.& 0.2657& N.A.& 0.1937&N.A.\\
            &AdvDrop      & 0.1649& +1.35\%& 0.1750& +0.63\%& 0.0729& +1.96\%& 0.0514& +1.38\%& 0.2660& +0.11\%& 0.1941& +0.33\%\\
            &GALORE      & 0.1625& -0.12\%& 0.1745& +0.35\%& 0.0732& +2.38\%& 0.0541& +6.71\%& 0.2633& -0.90\%& 0.1927& -0.52\%\\
            &APDA       & \underline{0.1657}& \underline{+1.84\%}& \underline{0.1772}& \underline{+1.90\%}& 0.0742& +3.78\%& 0.0526& +3.75\%& \textbf{0.2679}& \textbf{+0.83\%}& \textbf{0.1957}&\textbf{+1.03\%}\\
            & AdjNorm     & 0.1628& +0.06\%& 0.1742& +0.17\%& 0.0739& +3.36\%& 0.0525& +3.55\%& 0.2626& -1.17\%& 0.1923&-0.72\%\\
            & Reg         & 0.1626& -0.06\%& 0.1742& +0.17\%& 0.0712& -0.42\%& 0.0503& -0.79\%& \underline{0.2664}& \underline{+0.26\%}& \underline{0.1948}&\underline{+0.57\%}\\
            & IPS         & 0.1654& +1.66\%& 0.1764& +1.44\%& \underline{0.0843}& \underline{+17.90\%}& \underline{0.0606}& \underline{+19.53\%}& 0.2474& -6.89\%& 0.1856&-4.18\%\\
            & MACR         & 0.1620& -0.43\%& 0.1733& -0.35\%& 0.0724& +1.26\%& 0.0513& +1.18\%& 0.2627& -1.13\%& 0.1921&-0.83\%\\
            & CAGED & \textbf{0.1677}& \textbf{+3.07\%}*& \textbf{0.1790}& \textbf{+2.93\%}*& \textbf{0.0857}& \textbf{+19.86\%}*& \textbf{0.0609}& \textbf{20.12\%}*& 0.2604& -1.99\%*& 0.1935&-0.10\%*\\
            \midrule
            \multirow{8}*{Epinions}& LightGCN    & 0.1239& N.A.& 0.1656& N.A.& 0.0241& N.A.& 0.0187& N.A.& 0.1883& N.A.& 0.1718&N.A.\\
            &AdvDrop      & 0.1242& +0.24\%& 0.1662& +0.36\%& 0.0249& +3.32\%& 0.0191& +2.14\%& 0.1871& -0.64\%& 0.1702& -0.93\%\\
            &GALORE      & 0.1247& +0.65\%& 0.1666& +0.60\%& 0.0253& +4.98\%& 0.0197& +5.35\%& 0.1868& -0.80\%& 0.1714& -0.23\%\\
            & APDA       & 0.1245& +0.48\%& 0.1661& +0.30\%& \underline{0.0260}& \underline{+7.88\%}& 0.0199& +6.42\%& \textbf{0.1933}& \textbf{+2.66\%}& \textbf{0.1743}&\textbf{+1.46\%}\\
            & AdjNorm     & 0.1250& +0.89\%& 0.1660& +0.24\%& 0.0248& +2.90\%& \underline{0.0203}& \underline{+8.56\%}& 0.1886& +0.16\%& 0.1709&-0.52\%\\
            & Reg         & 0.1234& -0.40\%& 0.1655& -0.06\%& 0.0244& +1.24\%& 0.0185& -1.07\%& 0.1866& -0.90\%& 0.1722&+0.23\%\\
            & IPS         & \underline{0.1253}& \underline{+1.13\%}& \underline{0.1664}& \underline{+0.48\%}& 0.0253& +4.98\%& 0.0195& +4.28\%& 0.1914& +1.65\%& 0.1746&+1.63\%\\
            & MACR         & 0.1230& -0.73\%& 0.1641& -0.91\%& 0.0240& -0.41\%& 0.0189& +1.07\%& 0.1874& -0.48\%& 0.1711&-0.41\%\\
            & CAGED & \textbf{0.1257}& \textbf{+1.45\%*}& \textbf{0.1674}& \textbf{+1.09\%*}& \textbf{0.0288}& \textbf{+19.50\%*}& \textbf{0.0226}& \textbf{+20.86\%*}& \underline{0.1916}& \underline{+1.75\%*}& \underline{0.1740}&\underline{+1.28\%*}\\
        \bottomrule
        \end{tabular}
    }
\end{table*}

\subsection{Baselines}
We select several representative baseline methods within the scope of popularity debiasing for performance comparison.
{\new These methods include both graph-based (edge reconstruction and aggregation re-weighting) and model-agnostic approaches (regularization, data valuation, and causal inference) to guarantee a comprehensive comparison.
Following previous studies \cite{galore, apda, advdrop}, we use \textbf{LightGCN} \cite{lightgcn}, which removes redundant activation units for efficient recommendation, as the backbone model.}
To maintain consistency and fairness, we fix the layer number as well as embedding dimensionality across all methods.
{\new Remarkably, we do not introduce more model-agnostic methods since we aim to develop graph-based debiasing solution instead of solely pursuing the state-of-the-art performance across all domains.}
The baselines are listed as follows. \\
Graph-based methods:
\begin{itemize}
    {\new \item \textbf{AdvDrop \cite{advdrop}:} employs adversarial edge dropout to enhance bias invariance in the interaction graph.
    \item \textbf{GALORE \cite{galore}:} strategically augmenting or dropping interaction edges based on the node importance.}
    \item \textbf{APDA \cite{apda}:} uses a normalized inner product between nodes to serve as the per-edge aggregation weight.
    \item \textbf{AdjNorm \cite{adjnorm}:} increases the power of normalization in aggregation weight.
    
\end{itemize}
Model-agnostic methods:
\begin{itemize}
    \item \textbf{Reg \cite{reg2}:} regularizes the correlation between item popularity and the corresponding inner product score in loss.
    \item \textbf{IPS \cite{ips2}:} weights training data to lower the importance of popular items in loss propagation.
    \item \textbf{MACR \cite{macr}:} applies causal counterfactual reasoning to relieve the popularity bias.
\end{itemize}

\subsection{Experiment Settings}
\subsubsection{\textbf{Data Preprocessing}}
We follow the experimental setup in \cite{galore} to binarize the explicit feedback in MovieLens, retaining ratings of 5 and discarding all other ratings. 
We split datasets into training set, validation set, and test set with a 7:1:2 ratio.
{\new For test set, we further set items with top 20\% node degree as popular items and the rest as niche items according to the Pareto Principle \cite{pareto}. }

\subsubsection{\textbf{Evaluation Metrics}}
To evaluate the model performances in Top-K recommendation, we employ two widely used metrics:
Recall@K measures the fraction of user's interested items that appear in the recommendations, while NDCG@K (Normalized Discounted Cumulative Gain) evaluates the ranking quality by putting more weights on the recommended items with higher rankings.
In the experiments, we set K $= 20$.

\subsubsection{\textbf{Hyper-parameter Settings}}
We implement CAGED and reproduce all the baselines in PyTorch. 
The LightGCN backbone is configured with 3 aggregation layers and 256-d embeddings. 
All models are initialized using the default normal initializer and optimized with the Adam \cite{adam} of 2048 batch size.
To better accommodate each dataset while ensuring fairness, LightGCN learning rate $\eta_1$ and regularization factor $\gamma$ vary across datasets but remain consistent in all models. Specifically, $\eta_1$ and $\gamma$ are tuned within $\{10^{-4}, 10^{-3}, 10^{-2}\}$.
For all the baselines, we follow the official reported hyper-parameter settings and perform grid search to fine-tune any ones lacking recommended values.
For CAGED, Two scaling factors $\lambda$ and $\beta$ are tuned in $[0, 1.4]$ with step $0.2$.
Mixture factor $\epsilon$ and CAGED learning rate $\eta_2$ are chosen from $\{10^{-4}, 10^{-3}, 10^{-2}\}$.

\subsection{Overall Performance (RQ1)}
To evaluate CAGED on Top-K recommendation and popularity debiasing, we comprehensively compare it with baselines in Table \ref{tab:main}, reporting Recall@20, NDCG@20, and improvement gains on all, niche, and popular items across datasets.
Results on all items reflect the overall recommendation performances, while those on niche items reflect the abilities of attention shift in these methods toward niche items, thereby indicating their debiasing effectivenesses.
We have the following observations.

Firstly, graph-based debiasing methods generally achieve improvements in recommendation accuracy as well as effective bias mitigation over the backbone model.
However, they consistently underperform CAGED in popularity debiasing, primarily due to the irrationalities in integrating graph aggregation.
Additionally, certain methods such as GALORE and APDA in some cases demonstrate marginal or even negative improvements on all items recommendation, indicating a decline in robustness caused by their irrationalities in balancing training and debiasing.

Secondly, among model-agnostic approaches, only IPS achieves both improved recommendation accuracy and effective debiasing.
In contrast, MACR reduces popularity bias at the cost of accuracy, and Reg achieves neither.
The potential reason is that model-agnostic methods only operate externally, lacking access to the underlying graph structure.
Thus, they have limited impact on mitigating the inherent popularity amplification in graph aggregation.

Lastly, CAGED delivers optimal improvements in both all items and niche items recommendations.
In terms of overall recommendation accuracy, CAGED achieves the improvements of $1.45\% \sim 3.07\%$ in recall and $1.02\% \sim 2.93\%$ in NDCG over the backbone model.
For cold items recommendation, it attains improvements of $19.50\% \sim 22.50\%$ in recall and $20.12\% \sim 22.81\%$ in NDCG.
With only minimal compromise in recommending popular items, CAGED shifts model focus toward niche items to greatly enhance both overall recommendation and debiasing qualities.
By effectively addressing the aforementioned irrationalities, CAGED demonstrates solid superiority over the above baselines.

\subsection{Study of Debiasing Functionality (RQ2)}
To illustrate how CAGED performs effective popularity debiasing to boost the recommendation performance, we analyze the evolution of inverse item popularity (IIP) throughout the weight update process.
We define IIP alternatively as the reciprocal of the square root item degree, which can be directly calculated by removing the square root of user degree from the original aggregation weight. 
According to the definition, popular items cluster near IIP = 0, while niche items gather around IIP = 1.

Figure \ref{fig:density} presents the IIP density estimations at different training stages, where all datasets exhibit consistent patterns.
Before the update, the densities are sharply concentrated near 0, with only a sparse distribution of niche items above 0.5, reflecting strong popularity bias.
During the update, the densities become more uniform and shift to higher IIPs.
Concurrently, the distributions of niche items smooth toward lower IIP regions, indicating that the model starts to balance its focus between popular and niche items.
After the update, the densities become notably less skewed toward 0, demonstrating a more balanced popularity distribution.

These findings confirm that CAGED effectively neutralizes the popularity imbalance between popular and niche items. 
By introducing the momentum update in aggregation weight, CAGED addresses the overemphasis of popular items in recommendations.

\begin{table}[t!]
    \centering
    \caption{Ablation study of CAGED training strategy.}
    \label{tab:ablation}
    \resizebox{\linewidth}{!}{
        \begin{tabular}{c |c c|c c|c c}
            \toprule
            \multirow{2}*{Variant} & \multicolumn{2}{c|}{MovieLens} & \multicolumn{2}{c|}{Pinterest} & \multicolumn{2}{c}{Epinions} \\
            ~  & R@20 & N@20 & R@20 & N@20 & R@20 & N@20 \\
            \midrule
            \multirow{2}*{\textsl{w/o-TS}} 
                & 0.2760 & 0.4807 & 0.1667 & 0.1771 & 0.1245 & 0.1648 \\
                & \textit{\textcolor{blue}{-1.43\%}} & \textit{\textcolor{blue}{-0.93\%}}
                & \textit{\textcolor{blue}{-0.60\%}} & \textit{\textcolor{blue}{-1.06\%}}
                & \textit{\textcolor{blue}{-0.95\%}} & \textit{\textcolor{blue}{-1.55\%}} \\
            \midrule[0.1pt]
            \multirow{2}*{\textsl{w/o-UC}} 
                & 0.2783 & 0.4819 & 0.1672 & 0.1787 & 0.1225 & 0.1629 \\
                & \textit{\textcolor{blue}{-0.61\%}} & \textit{\textcolor{blue}{-0.68\%}}
                & \textit{\textcolor{blue}{-0.30\%}} & \textit{\textcolor{blue}{-0.17\%}}
                & \textit{\textcolor{blue}{-2.55\%}} & \textit{\textcolor{blue}{-2.69\%}} \\
            \midrule[0.1pt]
            \multirow{2}*{\textsl{w/o-MU}}
                & 0.1147 & 0.2782 & 0.0809 & 0.1014 & 0.0232 & 0.0517 \\
                & \textit{\textcolor{blue}{-59.04\%}} & \textit{\textcolor{blue}{-42.66\%}}
                & \textit{\textcolor{blue}{-51.76\%}} & \textit{\textcolor{blue}{-43.35\%}}
                & \textit{\textcolor{blue}{-81.54\%}} & \textit{\textcolor{blue}{-69.12\%}} \\
            \midrule[0.1pt]
            \textbf{CAGED}  
                & \textbf{0.2800} & \textbf{0.4852} 
                & \textbf{0.1677} & \textbf{0.1790} 
                & \textbf{0.1257} & \textbf{0.1674} \\
            \bottomrule
        \end{tabular}
    }
\end{table}

\begin{figure*}[t!]
    \centering  
    \subfigcapskip=-3pt 
    \subfigure[MovieLens]{\includegraphics[width=0.32\linewidth]{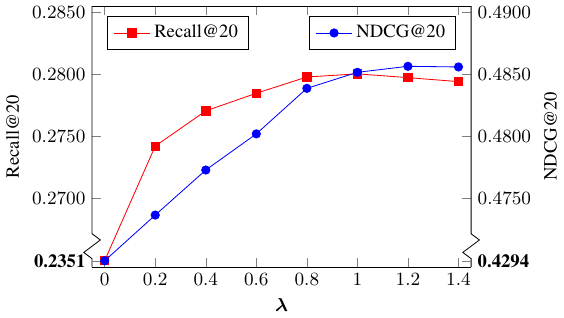}}
    \hspace{0.21cm}
    \subfigure[Pinterest]{\includegraphics[width=0.32\linewidth]{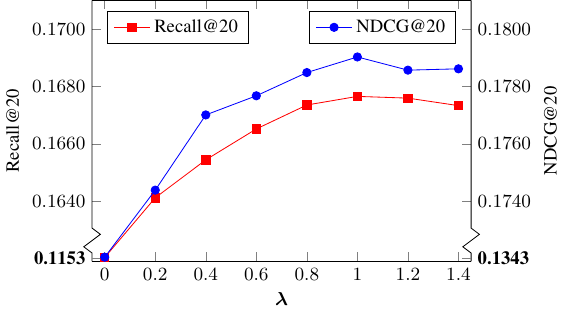}}
    \hspace{0.21cm}
    \subfigure[Epinions]{\includegraphics[width=0.32\linewidth]{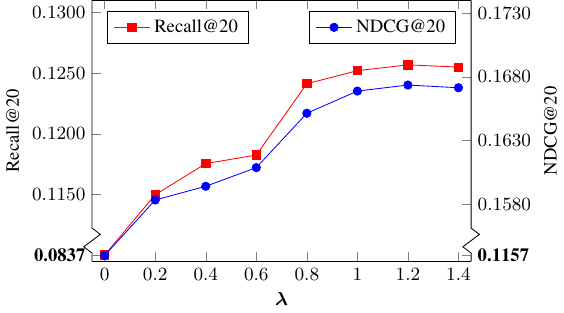}}
    \Description{lambda experiment}
    \caption{Recommendation performances of CAGED under different $\boldsymbol{\lambda}$s.}
    \label{fig:lam}
\end{figure*}

\begin{figure*}[t!]
    \centering  
    \subfigcapskip=-3pt 
    \subfigure[MovieLens]{\includegraphics[width=0.32\linewidth]{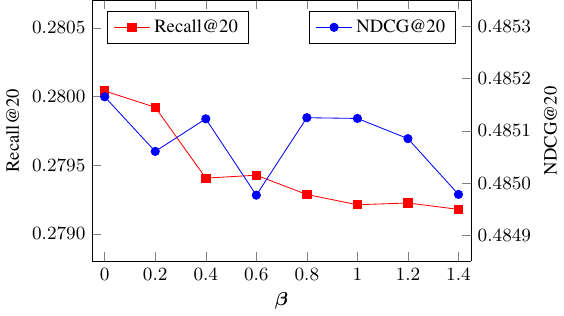}}
    \hspace{0.21cm}
    \subfigure[Pinterest]{\includegraphics[width=0.32\linewidth]{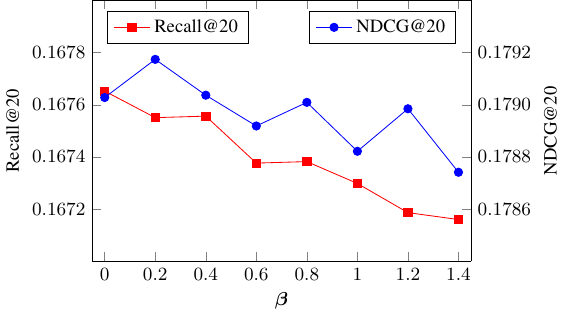}}
    \hspace{0.21cm}
    \subfigure[Epinions]{\includegraphics[width=0.32\linewidth]{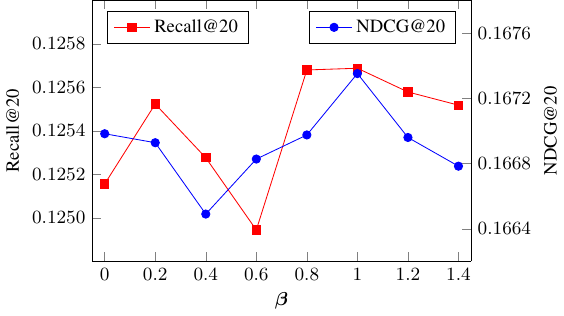}}
    \Description{beta experiment}
    \caption{Recommendation performances of CAGED under different $\boldsymbol{\beta}$s.}
    \label{fig:beta}
\end{figure*}

\subsection{Study of CAGED Training Strategy (RQ3)}
{\new To verify the effectiveness of each mechanism in our training strategy, we conduct the ablation study of three variants as shown in Table \ref{tab:ablation}}.
All variants show performance degradations, confirming the significant contribution of each mechanism to the performance.
Below, we analyze each variant in detail.
\begin{itemize}
    \item \textbf{w/o-TS:} 
    this variant removes the two-stage training strategy by pretraining CAGED on fixed LightGCN embeddings and then proceeding without further fine-tuning. 
    Performance degradation in this variant occurs because CAGED is restricted to be pretrained only on deterministic embeddings, keeping it from learning higher-quality embeddings.
    In contrast, two-stage training enables iterative refinement, leading to superior performance.
    
    \item \textbf{w/o-UC:} this variant removes the update condition to execute training and update without checking whether the recommendation performance improves at each epoch.
    This variant lowers the performance because CAGED suffers from the impact of under-trained representations.
    Hence, this condition ensures that CAGED only takes high-quality embedding as input, as the way to enhance overall performance.

    \item \textbf{w/o-MU:} this variant removes the momentum update strategy and directly replaces the aggregation weight matrix with the one generated by CAGED.
    However, it totally fails to converge.
    Since LightGCN module is initialized using the adjacent normalization weights to ensure a high-quality start,
    this abrupt replacement causes instability and loss explosion, as CAGED-generated weights misalign with original LightGCN weights.
    Momentum update is thus essential, providing a smooth transition that enables effective adaptation and stable optimization.
\end{itemize}

\subsection{Study of CAGED Components (RQ4)}
{\new To evaluate how each CAGED component affect the model performance, we conduct the following analysis in two parts.}
First, we investigate the ELBO formulation by adjusting hyper-parameters $\lambda$ and $\beta$.
{\new Second, we examine the construction of encoder and decoder in terms of layer number.}
To simplify the demonstration, we employ the same MLP structures for them.

\subsubsection{\textbf{ELBO Formulation Analysis}}
{\new To analyze the performance contribution of ELBO reconstruction term, we conduct the experiment as shown in Figure \ref{fig:lam}.}
Different values of $\lambda$ represent varying integration ratios.
The results on all datasets demonstrates that removing the reconstruction term severely impairs the performance.
As $\lambda$ increases, both recall and NDCG improve, achieving optimal performance at $\lambda = 1$ or $1.2$. 
Since the reconstruction term generated by the decoder directly corresponds to the conditional likelihood target, discarding it destroys the model’s ability to learn the data likelihood. 
In addition, performance becomes stable as $\lambda$ increases beyond a certain point.
We conclude that an adequate ratio of the reconstruction term in ELBO enables the CAGED to effectively predict the aggregation weight target.

{\new To analyze the impact of incorporating the KL term into ELBO at different ratios, we conduct the experiment as shown in Figure \ref{fig:beta}.}
Remarkably, setting $\beta$ to 0 completely discards the KL term in ELBO and setting $\beta$ to 1.4 further amplifies it.
In MovieLens and Pinterest, we observe that omitting KL term results in the best Recall@20 performance, and increasing $\beta$ leads to performance degradation in both recall and NDCG.
However, for Epinions, $\beta = 1$ yields the optimal performance, and further increase or decrease in $\beta$ causes a decline in recommendation result.
Given the nature that the KL term in ELBO plays as the role of regularization to prevent the latent space from diverging too far from its prior distribution, we conclude that different datasets exhibit varying sensitivities to this term generated by the encoder.
Therefore, this finding highlights the significance of carefully tunning $\beta$ for CAGED to adapt effectively to different training environments.

\begin{table}[t!]
    \centering
    \caption{Recommendation performances of CAGED under different encoder \& decoder layer number.}
    \label{tab:layer}
    \resizebox{\linewidth}{!}{\new
        \begin{tabular}{c |c c|c c|c c}
            \toprule
            \multirow{2}*{Variant} & \multicolumn{2}{c|}{MovieLens} & \multicolumn{2}{c|}{Pinterest} & \multicolumn{2}{c}{Epinions} \\
            ~  & R@20 & N@20 & R@20 & N@20 & R@20 & N@20 \\
            \midrule
            \multirow{2}*{\textsl{1-Layer}} 
                & 0.2787 & 0.4837 & 0.1665 & 0.1780 & 0.1244 & 0.1655 \\
                & \textit{\textcolor{blue}{-0.46\%}} & \textit{\textcolor{blue}{-0.31\%}}
                & \textit{\textcolor{blue}{-0.72\%}} & \textit{\textcolor{blue}{-0.56\%}}
                & \textit{\textcolor{blue}{-1.03\%}} & \textit{\textcolor{blue}{-1.14\%}} \\
            \midrule[0.1pt]
            \multirow{2}*{\textsl{2-Layers}} 
                & 0.2794 & 0.4849 & 0.1670 & 0.1784 & 0.1255 & 0.1671 \\
                & \textit{\textcolor{blue}{-0.21\%}} & \textit{\textcolor{blue}{-0.27\%}}
                & \textit{\textcolor{blue}{-0.42\%}} & \textit{\textcolor{blue}{-0.34\%}}
                & \textit{\textcolor{blue}{-0.16\%}} & \textit{\textcolor{blue}{-0.18\%}} \\
            \midrule[0.1pt]
            \multirow{2}*{\textsl{4-Layers}}
                & 0.2799 & 0.4845 & 0.1676 & 0.1792 & 0.1256 & 0.1672 \\
                & \textit{\textcolor{blue}{-0.03\%}} & \textit{\textcolor{blue}{-0.35\%}}
                & \textit{\textcolor{blue}{-0.06\%}} & \textit{\textcolor{gray}{+0.11\%}}
                & \textit{\textcolor{blue}{-0.08\%}} & \textit{\textcolor{blue}{-0.12\%}} \\
            \midrule[0.1pt]
            \textbf{3-Layers}  
                & \textbf{0.2800} & \textbf{0.4852} 
                & \textbf{0.1677} & \textbf{0.1790} 
                & \textbf{0.1257} & \textbf{0.1674} \\
            \bottomrule
        \end{tabular}
    }
\end{table}

\subsubsection{\textbf{\new Encoder \& Decoder Structure Analysis}}
To evaluate the impact of layer number, Table \ref{tab:layer} compares the performances using 1 to 4 neural network layers to construct the encoder and decoder.
In each configuration, we use the same number of neurons in input and output layer. 
For hidden layers, we follow the strategy of doubling dimension in each layer up to the midpoint, and symmetrically decreasing afterward.
From the results, we observe that networks with fewer than 3 layers perform worse, mainly due to their simplistic constructions.
The 3-layer network structure exhibits optimal performance in Recall@20 across all three datasets.
However, 4-layers fails to bring improvement in recall for all the datasets and shows a drop in NDCG for MovieLens and Epinions, which indicates overfitting potential under increasing model complexity.
Therefore, we adopt the 3-layer network structure for all datasets in the experiments, as it balances simplicity with stability to well-learn the interaction likelihood.

\section{Related Work}
Research on addressing popularity bias in recommender systems has a relatively long history, with numerous studies contributing from various perspectives \cite{traditionaldebiasing}. 
A simple approach is adding regularization in loss function to prevent models from overfitting to popular items \cite{reg1, reg2, reg3, popularitybias}. 
Another straightforward idea is data valuation, which controls the influence of popularity by directly weighting data nodes \cite{value}. 
For example, inverse propensity score, which is a fundamental metric for evaluating data likelihood, is utilized in \cite{ips1, ips2, ips3, closedloop} to weight training data for alleviating bias.
Beyond these, more works \cite{tailnet, advncf} focus on modifying CF model structure by integrating powerful auxiliary modules. 
{\new Under the growing interest in graph-based recommendations, graph-based debiasing has emerged as a promising research direction in parallel to the above model-agnostic methods.}
For example, APDA \cite{apda} refines aggregation weights by proposing a novel metric to measure node popularity.
{\new AdjNorm \cite{adjnorm} regularizes aggregation weights to inhibit popularity.}
GALORE \cite{galore} and AdvDrop \cite{advdrop} address popularity bias by strategically adding or removing graph edges.
However, as discussed earlier, these graph-based approaches have limitations and remain inadequate in fully addressing popularity bias.

The introduction of causal inference into recommender systems offers a different perspective of popularity debiasing.
Its fundamental idea is to theoretically model bias within a causal graph and apply probabilistic adjustment to deconfound the bias.
For example, PDA \cite{pda} and CCF \cite{casualcf} introduce backdoor adjustment to block the confounding effect and utilize causal intervention to estimate user-item preferences.
The flexibility in causal graph design allows causal inference to address diverse confounding effects.
COR \cite{cor} targets out-of-distribution recommendation and DCCF \cite{dccf} models unobserved personalized confounding effects.
A worth mentioning study, NCGCF \cite{ncgcf} reconstructs graph aggregation to capture causal information, however, we leave this work out as it explores another direction distinct from tackling popularity bias.
In general, these methods have not focused on taking the advantage of causal inference to reflect graph-based recommendation rationale.
Additionally, some works are restricted to customized causal graphs rather than fitting a general scenario.

\section{Conclusion and Future Work}
In this paper, we present the CAGED model to tackle popularity bias in graph-based recommendations.
First, we utilize causal inference to formalize graph aggregation as a form of backdoor adjustment.
Next, we implement our model to learn the aggregation weights by approximating the unbiased history likelihood.
Combined with the proposed momentum update strategy, we address two common irrationalities in existing graph debiasing methods.
Extensive experiments verify the superior debiasing effectiveness of CAGED.
Our future work will investigate the integration of distinct likelihood estimation into each aggregation layer to capture the popularity bias more effectively, given that GCN operates in a layer-wise manner.

\begin{acks}
This work was partially supported by the Early Career Scheme (No. CityU 21219323) and the General Research Fund (No. CityU 11220324) of the University Grants Committee (UGC), the NSFC Young Scientists Fund (No. 9240127), Hong Kong Research Grants Council's Research Impact Fund (No.R1015-23), Collaborative Research Fund (No.C1043-24GF), General Research Fund (No.11218325), Institute of Digital Medicine of City University of Hong Kong (No.9229503), Huawei (Huawei Innovation Research Program), Tencent (CCF-Tencent Open Fund, Tencent Rhino-Bird Focused Research Program), Alibaba (CCF-Alimama Tech Kangaroo Fund No. 2024002), Ant Group (CCF-Ant Research Fund), Didi (CCF-Didi Gaia Scholars Research Fund), Kuaishou, Bytedance, and the Graduate Research Fund of the School of Economics and Management of Dalian University of Technology (No. DUTSEMDRFKO1).
\end{acks}

\section*{GenAI Usage Disclosure}
We only use the GPT-4.1 tool in manuscript writing to improve the grammar and clarity of existing text.
At other research stages including code implementation and experiments, no GenAI tools were used.
We take full responsibility for the content presented.

\balance

\bibliographystyle{ACM-Reference-Format}
\bibliography{ref}
\end{document}